\documentclass[aps,pra,twocolumn,groupedaddress,showpacs,]{revtex4-1}
\usepackage{latexsym}
\usepackage{graphicx}
\usepackage{amsmath} 
\usepackage{amssymb} 
\usepackage{color}
\def\eqnref#1{Eq.~(\ref{#1})}
\def\figref#1{Fig.~(\ref{#1})}

\begin{document}

\title{Electron swarm experiments in dense rare gases: a review.}
\author{A. F. Borghesani} \email{E-mail  address: armandofrancesco.borghesani@unipd.it}
 \affiliation{CNISM Unit, Department of Physics \& Astronomy\\ University of Padua, Padua, Italy }
\begin{abstract}
Swarm techniques have largely been used to investigate electron transport in very dilute gases in order to shed light on the electron-atom (molecule) scattering cross section and, hence, on the interaction potential. The theoretical basis for the analysis of these experiments is classical Kinetic Theory. 
However, electron transport in dense media, either in gaseous- or condensed phase, is a physical phenomenon of fundamental and practical interest. Dense rare gases are model systems for disordered media. They are particularly well suited to investigate how the dynamics and energetics of quasifree electrons change as the environment density is gradually increased. A review on the electron swarm experiments in dense rare gases is presented here. 
\end{abstract} 
\maketitle
\section{Introduction}\label{sec:intro}
The investigation of the transport properties of quasifree electrons in dilute gases has been pursued for many years now 
not only because it gives insight into the electron-atom or electron-molecule interaction by directly measuring the scattering cross sections (either differential, momentum transfer, or total)~\cite{huxcro} but also because many applications rely on electronic conduction. However, many of such applications, for instance liquid ionization chambers or medical imaging systems among others, are based on electron conduction in liquids. In {\em fast} liquids, i.e., those in which the electrons are very mobile and their mean free path (mfp) is very long, the transport properties of charge carriers can still be described within the frame of reference of classical Kinetic Theory (cKT) provided that the scattering cross sections are considered as parameters that depend on the liquid density and structure~\cite{lekner1967,vanhove1954}, including density fluctuations~\cite{atrazhev2002} and phonon spectrum~\cite{naveh1993}. 

Classical Kinetic Theory implements a single-scat\-te\-ring picture even in the liquids although the fundamental hypotheses for its validity are not fulfilled. The study of electron transport in dense gases allows to shed light how cKT still retains its validity far beyond expectation. 

Dense rare gases are chosen to investigate how the dynamics and energetics of electrons are modified as the environment density is changed on the basis of several reductionist reasons. They are clean systems with large ionization energies so that the amount of charge carriers in the experiment is readily controlled by scientists. Their equations of state are very well known so that their density can accurately be computed if pressure and temperature are known. They are very good models of disordered systems. 

Over the years, the experimental results of electron swarm experiments in dense rare gases have ascertained the existence and importance of several multiple scattering (MS) effects that not only affect the transport properties of electrons but several other properties of electrons in dense nonpolar fluids. A proper account of MS have allowed researchers to recognize that the different phenomenology of the electron behavior in various gases can be attributed to the different shape of the scattering cross sections within a unique physical picture. A heuristic model has thus been established that incorporates the MS effects into the single scattering picture of cKT in such a way that the equations of cKT can be extended to a much wider density range than previously expected. 

In this paper a review of the experimental results on electron transport in dense rare gases is presented along with a discussion of the MS effects. It will be shown how the equations of cKT are heuristically modified so as to properly describe MS and how a single scattering picture can be retained. Moreover, some related physical phenomena observed in electron swarm experiment at high density are presented and discussed.

\section{Electron Transport\label{sect:etransport}}
In a typical swarm experiment a bunch (or beam) of electrons is drifted through the gas under the action of an externally applied electric field $E.$ The  transit time between cathode and anode is recorded and the electron drift velocity can be  easily determined. Under the typical conditions in dense gases, diffusion can be neglected and the drift mobility is determined.
For not too strong electric fields $E$, the two-term approximation of the solution to the Boltzmann equation is valid and cKT gives an explicit formula for 
the electron drift mobility $\mu,$ provided that only elastic scattering processes occur~\cite{huxcro}
\begin{equation}\label{eq:munCL}
\mu N = -\frac{e}{3}\left(\frac{2}{m}\right)^{1/2}\int\limits\frac{\epsilon}{\sigma_\mathrm{mt}(\epsilon)}\frac{\mathrm{d}g(\epsilon)}{\mathrm{d}\epsilon}  
\,\mathrm{d}\epsilon
\end{equation}
$N$ is the number density of the gas, $e$ and $m$ the electron charge and mass, respectively, $\epsilon$ is the electron energy, and $\sigma_\mathrm{mt}(\epsilon)$ is the 
electron-atom mo\-men\-tum tran\-sfer scat\-te\-ring cross sec\-tion. $g$ is the Davydov-Pidduck distribution function~\cite{lekner1967,wannier}
\begin{eqnarray}
& & g(\epsilon)= A\times  \hfill  \nonumber \\
&& \exp{\left\{-
\int\limits_0^\epsilon 
\left[
k_\mathrm{B}T+\frac{M}{6m}\left(\frac{eE}{N}\right)^2\frac{1}{z\sigma^2_\mathrm{mt}(z)}
\right]^{-1}
\,\mathrm{d}z
\right\}
}
\label{eq:davpid}
\end{eqnarray}
\noindent 
$k_\mathrm{B}$ is the Boltzmann constant, $T$ the absolute temperature, and $M$ is the atom mass. $A$ is a normalization constant determined by $\int_0^\infty \epsilon^{1/2} g(\epsilon) \, \mathrm{d}\epsilon=1
.$
As a consequence of \eqnref{eq:munCL} and \eqnref{eq:davpid}, the density normalized mobility $\mu N$ for a given gas turns out to be solely a function of $T$ and of the reduced electric field $E/N.$
 
For $E/N\rightarrow 0,$ cKT predicts that the zero-field density normalized mobility $\mu_0 N$ is given by
\begin{equation}
\left(\mu_0 N\right)_\mathrm{cl} = \frac{4e}{3\left[2\pi m \left(k_\mathrm{B} T\right)^5 \right]^{1/2}}\int\limits_0^\infty\frac{\epsilon}{\sigma_\mathrm{mt}(\epsilon)}e^{-\epsilon/k_\mathrm{B}T}\,\mathrm{d}\epsilon
\label{eq:mu0ncl}
\end{equation}
$ \left(\mu_0 N\right)_\mathrm{cl}$ can be considered as a suitable thermal average of the inverse cross section and, for any given gas, it turns out to be only function of $T.$

\begin{figure}[t!]
\resizebox{1\columnwidth}{!}{%
\hskip 1.5 cm  \includegraphics{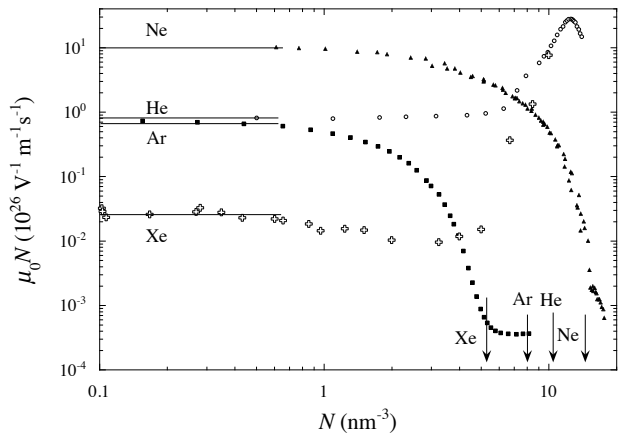}
}
\caption{$\mu_0 N$ vs $N.$ Triangles: Ne at $T= 45\,$K~\cite{borg1990b}. Circles: Ar at $T=152.2\,$K~\cite{borg2001}. Squares: He at $T=26\,$K~\cite{borg2002}. Crosses: Xe at SVP~\cite{huang1978}. 
Solid lines: prediction of cKT. Arrows: critical densities.\label{fig:mu0NHeNeArXe}}
\end{figure}

Experiments, however, strongly contradict the theoretical prediction. In \figref{fig:mu0NHeNeArXe}
the experimental results for $\mu_0 N$ are shown as a function of $N$ at relatively low $T$ for  Ne~\cite{borg1990b}, Ar~\cite{borg2001}, He~\cite{borg2002}, and Xe~\cite{huang1978}. Except Xe, in which the measurements have been carried out under saturated vapor pressure (SVP) conditions, all other experiments have been carried in the one-phase region along isotherms.
The solid lines on the left show the constant value $\mu_0 N$ should have according to cKT. The arrows at the bottom of the figure indicate the critical density $N_{c}$ of each gas. 

All gases show pronounced deviations from the prediction of cKT that have been observed long ago (for a review, see Ref.~\cite{Christophourou1984a,christophorou1984b}). $\mu_0 N$  strongly decreases with increasing $N$ ({\it negative density effect}) in He and Ne, whose electron-atom interaction is mainly repulsive because of the short-range exchange forces leading to a positive value of their scattering length $a$. 
By contrast, in negative scattering length gases, such as Ar and the heavier noble gases, in which the dominant interaction is due to long-range polarization forces, $\mu_0 N$ increases with increasing $N$ ({\it positive density effect}). 

The evident failure of cKT is due to the fact that the conditions which it is based on are no longer valid. At the density and temperatures of the experiments the mfp and the electron thermal wavelength $\lambda_T=h/\sqrt{2\pi mk_\mathrm{B}T}$ become comparable to each other and both to the average interatomic distance $d\sim N^{-1/3}$. In such circumstances, MS effects are no longer negligible and have to be taken into account.

Historically, the two density effects shown by $\mu_0 N$ have theoretically been treated in separate ways (for a review, see Ref.~\cite{braglia1982}). The positive density effect was mainly attributed to a shift of the kinetic energy of the quasifree electrons due to MS, whereas the negative density effect was mainly attributed to a quantum self-interference of the electron wave packet that leads to an increase of the scattering rate~\cite{atrazhev1977,polishuk1984}.

\begin{figure}[t!]
\resizebox{1\columnwidth}{!}{%
  \includegraphics{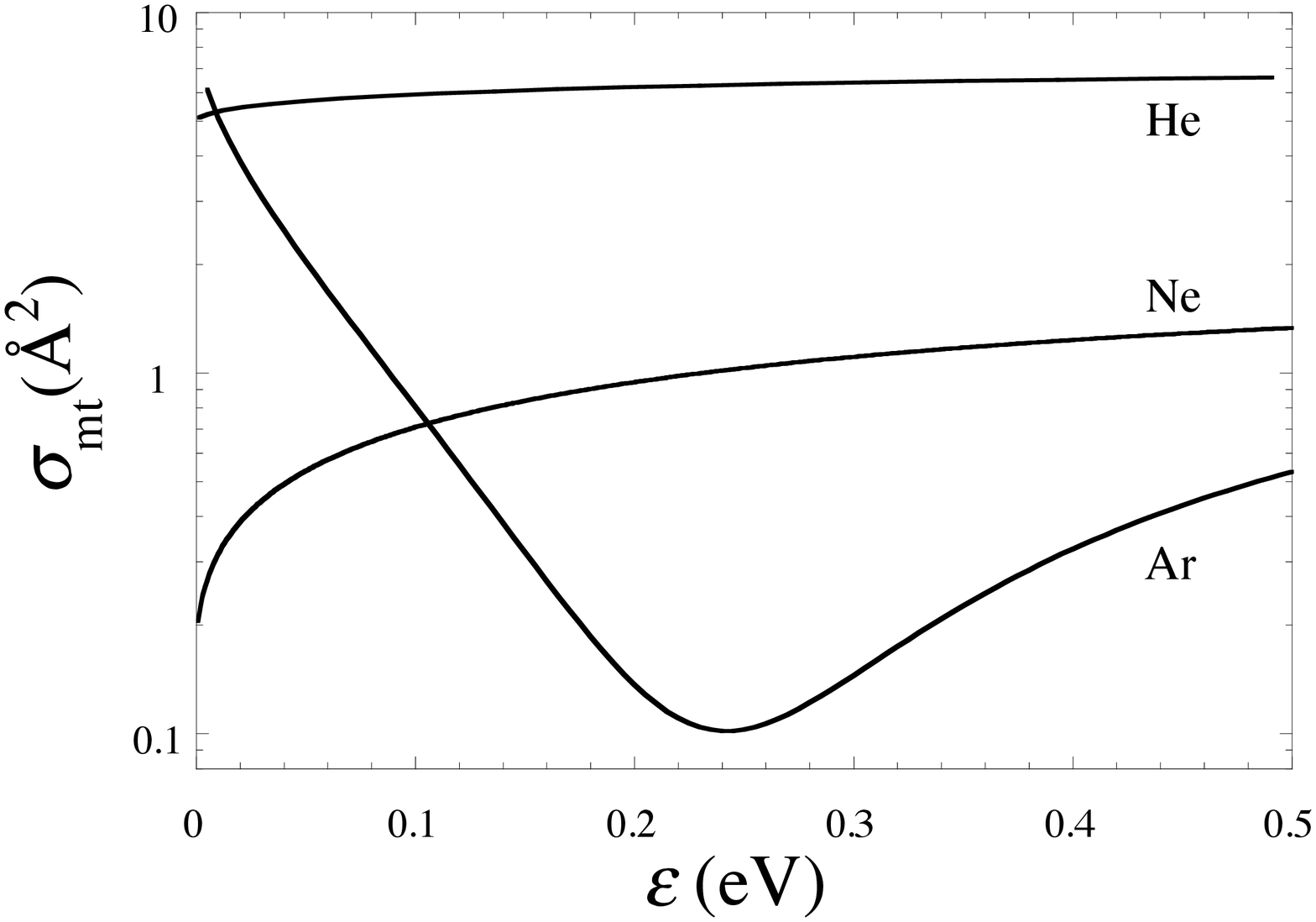} 
}\caption{\small $\sigma_\mathrm{mt}$ vs $\epsilon$ for He~\cite{omalley1963}, Ne~\cite{omalley1980}, and Ar~\cite{weyhreter1988}. \label{fig:sezurtArHeNe}}
\end{figure}

It is now clear that MS affects the behavior of quasifree electrons in all gases in the same way. However, the experimental outcome depends on the energy dependence of the electron-atom momentum transfer scattering cross section $\sigma_\mathrm{tm}(\epsilon)$, some of which are shown in \figref{fig:sezurtArHeNe}.

Three main MS effects have been highlighted~\cite{BSL}. 
The first one is a density dependent quantum shift $V_0 (N) $ of the electron energy~\cite{fermi1934}. The mean energy of an electron in thermal equilibrium with the gas is enhanced to $\langle \epsilon\rangle=(3/2) k_\mathrm{B}T +V_0(N).$ $V_0(N)$ consists of two contributions~\cite{springett1968}
\begin{equation}\label{eq:v0n}
V_0 (N)=U_p(N)+E_k(N) \end{equation}
$U_p<0$ is a potential energy term due to the polarization induced by the electron in the surrounding gas. 
$E_k>0$ is a kinetic energy contribution stemming from the shrinking of the volume available to the electron upon increasing $N.$ 
$E_k$ is obtained in terms of the wavevector $k_0(N)$ as
\begin{equation}\label{eq:ek}
E_k (N)= \frac{\hbar k_0^2}{2m}\end{equation}
by enforcing local translational invariance of the electron wave packet over the diameter $2r_s$ of the Wigner-Seitz sphere~\cite{hernandez1991} yielding the eigenvalue equation
\begin{equation}\label{eq:k0}
k_0 r_s =\tan{\left[k_0\left(r_s -\tilde a\right)\right]}\end{equation}
with $(4\pi r_s^3/3)N=1.$  $\tilde a=\sqrt{(\sigma_T/4\pi)}$ is the Hartree-Fock hard-sphere radius of the atom and $\sigma_T$ is the total scattering cross section~\cite{miyakawa1969}. 

It has been shown~\cite{BSL} that the dynamic properties of the electrons are affected by only $E_k.$ The electron kinetic energy at collision is $\epsilon= p^2/2m +E_k(N),$ whereas it is the group velocity $v=\sqrt{(2/m)(\epsilon -E_k)}$ that contributes to the energy equipartition value arising from the gas temperature.

The relevance of this first MS effect depends on how $\sigma_\mathrm{mt}$ depends on energy.
According to \eqnref{eq:mu0ncl},  $\mu_0 N$ is a kind of thermal average of $(1/\sigma_\mathrm{mt}).$
Upon increasing $N$, $E_k(N)$ and the mean electron energy increase as well. This fact corresponds to evaluating the inverse cross section at increasing energy. It is clear, by inspecting \figref{fig:sezurtArHeNe}, that an increase of the mean energy in Ar leads to a decrease of the average cross section and, hence, to an increase of $\mu_0 N$ because $\sigma_\mathrm{mt}$ rapidly decreases with increasing energy. The qualitatively similar energy dependence of the cross sections of the heavier gases Kr and Xe~\cite{frost1964} is mirrored by the similar behavior of $\mu_0N.$

On the contrary, the opposite effect occurs in Ne, for which the cross section rapidly increases with energy. 
Finally,  $\sigma_\mathrm{mt}$ in He is nearly constant and the quantum energy shift does not influence $\mu_0 N$ very much. 

The second MS effect is a quantum self-interference of the electron wave packet that propagates along paths connected by time reversal symmetry~\cite{atrazhev1977,polishuk1984,ascarelli1986,adams1992}.
This self-interference leads to a fractional increase of the cross section that is a function of the ratio of the electron wavelength $\lambda$ to its mfp $\ell.$ This effect is large when the cross section is large. It is particularly important for He, whose cross section is large and nearly energy in\-de\-pen\-dent, whe\-re\-as it is less effective in Ne, whose cross section is small, and is mildly important in Ar, whose cross section is large at low energy but rapidly decreases with increaing $\epsilon.$ In He, the effect becomes so large as to lead to the appearance (at very large $N$) of a mobility edge and to the phenomenon of weak localization~\cite{polishuk1984}.

The third MS effect is important close to the critical point, where the correlations among scatterers are strong. The scattered wave function is obtained by coherently summing up the partial amplitudes scattered by each individual atom. By so doing, the cross section is enhanced by the static structure factor of the gas, $S(k),$ whose long wavelength limit $S(0)$ is related to the gas isothermal compressibility $\chi_T$ as $S(0)=Nk_\mathrm{B}T\chi_T$~\cite{lekner1968,lekner1973}. 

The acknowledgment of the influence of MS has led to the proposal of a heuristic  model~\cite{BSL} that incorporates the MS concepts into the single scattering picture of cKT. The basic idea is that MS modifies the ``bare'' cross sections yielding density dependent ``effective'' cross sections. The procedure to obtain this goal is valid for all gases and does not introduce any freely adjustable parameters. 

According to the heuristic model, the equations of cKT \eqnref{eq:munCL} and \eqnref{eq:davpid} are retained provided that the cross section $\sigma_\mathrm{mt}$ is replaced by the effective cross section $\sigma_\mathrm{mt}^\star $ given by
\begin{equation}
\sigma_\mathrm{mt}^\star (w,N) =F(w)\sigma_\mathrm{mt}(w)\left[
1+
\frac{2\hbar F(w)N\sigma_\mathrm{mt}(w)}{\left(
2mw
\right)^{1/2}}
\right]
\label{eq:sigmastar}
\end{equation}
in which $w=\epsilon +E_k(N)$ is the energy shifted by the density dependent quantum shift $E_k(N).$
$F$ is the correlation factor~\cite{lekner1968}
\begin{equation}\label{eq:F}
F(k)=\frac{1}{4k^2}\int\limits_0^{2k}q^3 S(q)\,\mathrm{d}q 
\end{equation} in which $k=\sqrt{2m\epsilon}$ is the usual relationship between energy and wave vector.
For not too large values of exchanged momentum $S(q)$ can be written as~\cite{stanley}
\begin{equation}\label{eq:sq}
S(q)=\frac{\left[
S(0)+ (qL)^2
\right]}{\left[1+ (qL)^2
\right]}
\end{equation}
with $L^2=0.1 l^2[S(0)-1]$ and $l\approx 10\,$\AA\ is the short-range correlation length.

According to Ref.~\cite{atrazhev1977}, to first order in $\lambda/\ell 
,$ the 
 weak localization correction due to quantum self-in\-ter\-fe\-ren\-ce enhances the cross section 
by the factor $(1+N\lambda\sigma_\mathrm{mt}/\pi)$ that appears in the square bracket in \eqnref{eq:sigmastar}. 

By so doing, \eqnref{eq:sigmastar} includes all the three MS effects: quantum shift of the electron energy, quantum self-interference, and scatterers correlation and, at the same time, the single scattering approach of cKT is preserved.

\begin{figure}[t!]
\resizebox{1\columnwidth}{!}{%
  \includegraphics{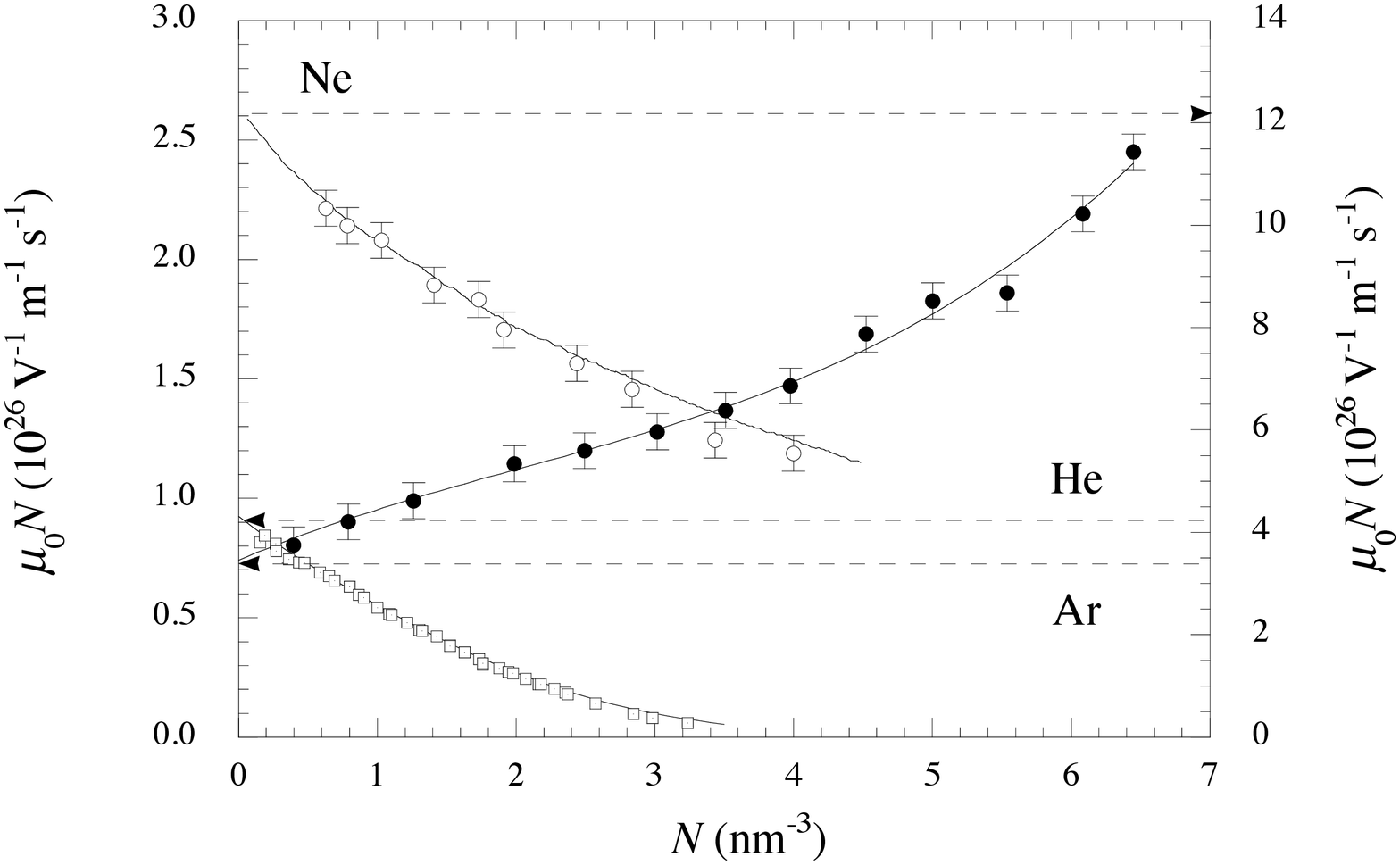}
}
\caption{$\mu_0 N$ vs $N$ up to intermediate $N.$ Open circles: Ne at $T=47.9\,$K (right scale)~\cite{borg1988}. Closed circles: Ar at $T=162.7\,$K (left scale)~\cite{BSL}. Open squares: He at $T=26.1\,$K (left scale)~\cite{borg2002}. Solid lines: heuristic model. Dashed lines: cKT.\label{fig:mu0NHeNeArLowN}}
\end{figure}

The heuristic model accurately describes the experimental results for $\mu_0 N$ up to quite large $N$, as shown in  \figref{fig:mu0NHeNeArLowN}. It is worth stressing the fact that no adjustable parameters are required. The differences between the several gases enter the model through the proper cross sections and equations of state.

\begin{figure}[b!]
\resizebox{1\columnwidth}{!}{%
 \hglue 0 cm \includegraphics{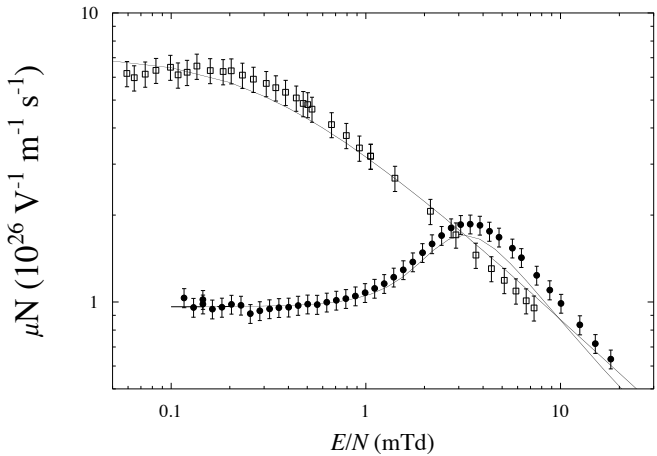}
}\caption{$\mu N$ vs $E/N$ at high density. Squares: Neon at $T=46.5\,$K and $N=1.89\,$nm$^{-3} $~\cite{borg1990b}. Circles: Argon at $T=152.15\,$K and $N=5.14\,$nm$^{-3}$~\cite{borg2001}. Lines: heuristic model.\label{fig:muNENnear}}\end{figure}

The heuristic model is also able to correctly describe the field dependence of $\mu N.$ As an example, $\mu N$ is plotted vs $E/N$ in \figref{fig:muNENnear} in Ne at $T=46.5\,$K and $N=1.89\,$nm$^{-3}$~\cite{borg1990b} and in Ar at $T=152.15\,$K for a density  $N=5.14\,$nm$^{-3}.$

\noindent It is to be noted that the validity of the prediction of the heuristic model can be extended to densities that are a significant fraction of the critical densities or even larger than that. For instance, in the case of Ar, the heuristic model is in agreement with the experiment up to $N=10\,$nm$^{-3}$~\cite{borg2001} whereas the critical density is $N_c=8.08\,$nm$^{-3}.$ On the contrary, in negative density effect gases the agreement of the heuristic model cannot be pushed to very large densities because of the onset of a different phenomenon, electron localization, which deserve a different treatment. In any case, the validity of the description of the heuristic model is extended up to $N\approx 10\,$nm$^{-3}<N_c=14.4\,$nm$^{-3}$ in Ne~\cite{borg1990b} and up $N\approx 3\,$nm$^{-3} <N_c=10.44\,$nm$^{-3}$ in He~\cite{borg2002}.

\begin{figure}[t!]
\resizebox{1\columnwidth}{!}{%
  \includegraphics{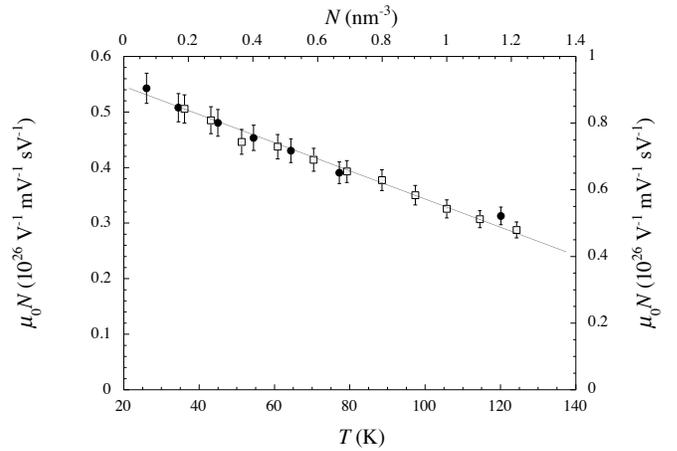}
}
\caption{Circles (left and bottom scales): $\mu_0N$ vs $T$ in He for $N\approx 1\,$nm$^{-3}$~\cite{neriphd}. Squares (right and top scales): $\mu_0 N$ vs $N$ for $T=26.1\,$K. Line: heuristic model.
\label{fig:V0heatsHe}}\end{figure}

\begin{figure}[b!]
\resizebox{1\columnwidth}{!}{%
  \includegraphics{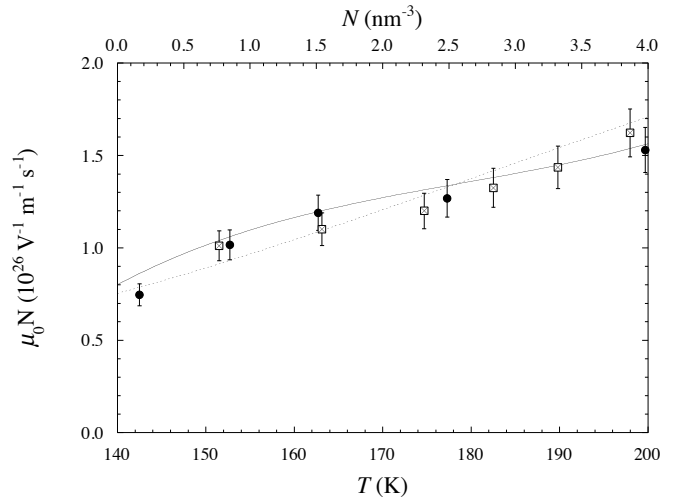}
}
\caption{Circles (and solid line): $\mu_0 N$ vs $N$ at constant in Ar $T=177.3\,$K (top scale)~\cite{borglamp}. Squares (and dotted line): $\mu_0 N$ vs $T$ at constant density $N=2.55\,$nm$^{-3}$ (bottom scale)~\cite{borglamp}. The lines are the prediction of the heuristic model for the two different thermodynamic conditions. 
\label{fig:V0heatsAr}}\end{figure}

There are more clues that the MS effect actually influence the dynamics and energetics of electrons in dense gases and there is also more evidence that the heuristic model correctly includes these effects in the cKT picture. For example, it interesting to note that the shift of the mean electron energy by the quantity $E_k (N)$ is equivalent to heating up electrons by increasing $N.$ This point of view is supported by \figref{fig:V0heatsHe} and \figref{fig:V0heatsAr} in which the dependence of $\mu_0 N$ as a function of $T$ at constant $N$ and as a function of $N$ at constant $T$ is shown
for He and Ar, respectively.
It can immediately observed that the same effect on the mobility, i.e., a decrease in He or an increase in Ar, can be obtained by either increasing $T$ or $N.$ Owing to presence of $E_k(N)$ that increases the mean electron energy, the action of $T$ and $N$ can somehow be interchanged.

 It is now clear that the transport properties of electrons in dense rare gases up to intermediate values of the density can be described within a unique frame of reference that is still based on cKT, in which elements of many-body physics, i.e., MS effects, are heuristically incorporated. In this way, the intuitive single-scattering picture of cKT is preserved. 

It is also important to stress the fact that the heuristic model provides a description of the mobility, which  is independent of the sign of the scattering length, thereby leading to a unique physical description of the phenomenon.

\section{Electron Localization\label{sect:loc}} 
In positive scattering length gases, owing to the strong repulsive electron-atom interaction, a quasifree electron is a state no longer favored in terms of free energy with respect to a state in which the electron is localized in an empty void if $N$ is sufficiently large and $T$ low enough~\cite{hernandez1991,khrapak1979}.  In this case, $E_k(N)$ may become larger than the isothermal work needed to expand a bubble to a finite radius and, thence, electrons become self-trapped in the bubble. Once trapped, the electron mobiity becomes extremely low and is limited by the hydrodynamic behavior of the bubble. This phenomenon was very well known in liquid (normal and superfluid) He (for a review, see Ref.~\cite{borg2007}) and was observed also in liquid Ne~\cite{bruschi1972b}.

In He gas at $T=4.2\,$K it was observed for the first time by Levine and Sanders~\cite{levine1967} and  later at higher temperatures~\cite{borg2002,harrison1973,schwarz1980}. It was also observed  in dense Ne gas close to the critical temperature~\cite{borg1990b}. 

\begin{figure}[b!]
 \resizebox{1\columnwidth}{!}{%
   \includegraphics{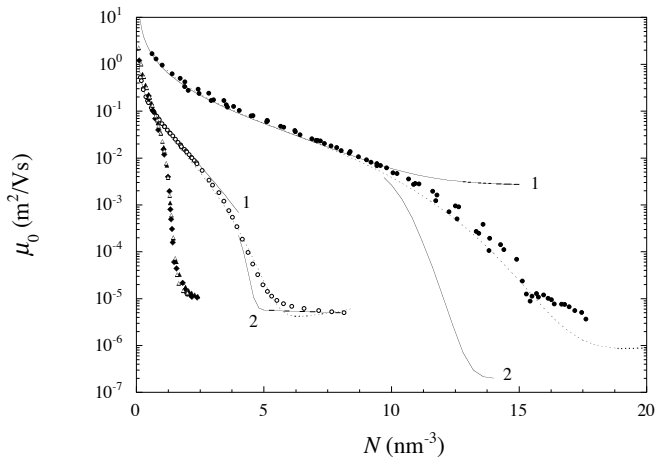}
   }
 \caption{$\mu_0$ vs $N.$ He at $T=4.2\,$K (diamonds~\cite{levine1967}, closed triangles~\cite{harrison1973}, and open triangles~\cite{schwarz1980}). Open circles: He at $T=26\,$K~\cite{borg2002}. Closed circles: Ne at $T\approx 46.5\,$K~\cite{borg1990b}. Solid curves 1: heuristic MS model for quasifree electrons~\cite{BSL}. Solid curves 2: optimum fluctuation model~\cite{borg1990b,borg2002}. Dashed curves: percolation model~\cite{borg1990b,eggarter1971}.
 \label{fig:MuvsNHeNeLocEgg}}\end{figure}

Phenomenologically, the localizaton manifests itself as a precipituous drop of the mobility as the density increases beyond a given threshold, as shown in~\figref{fig:MuvsNHeNeLocEgg}.
The localization phenomenon is not limited to low temperature. Actually, it has been observed in He gas for temperatures as high as $T\simeq 77\,$K~\cite{borg2002}. The dynamics of bubble formation is still not completely understood although it is believed to occur on the timescale of several hundreds of ps. Sakai {\em et al.}~\cite{sakai1992} argue that the electron undergoes a fast vertical transition to a virtual or resonant state in the continuum as soon as it crosses a less-than-average density fluctuations. Then, a slow, adiabatic process takes place in which the electron exerts a pressure on the wall of the fluctuation because of its repulsive interaction with the gas atoms. As the gas is compliant, this pressure pushes the gas atoms away thereby making the cavity larger and larger until an equilibrium condition is reached. Other researchers~\cite{eggarter1971} suggest that electrons might get self-localized because of strong disorder~\cite{anderson1958}. Once the incipient localization starts occurring, then the repulsive interaction and the gas compliance lead the process to its conclusion.

Calculations to describe the electron self-trapped state have been carried out in the so called {\em optimum fluctuation model} (OFM) by using an energy minimization process~\cite{fetter1975}. The free energy excess $\Delta F$ between the qua\-si\-free- and the localized state is computed as a function of $N,$ $T,$ and of the bubble radius $R.$ By minimizing $\Delta F$ with respect to $R$ the energy and radius of the most probable bubble state is obtained. There are several models to implement the calculation of $\Delta F$, some of them are based on a self-consistent field approximation~\cite{hernandez1991,khrapak1979,volykhin1995}, whereas some others are much simpler~\cite{miyakawa1969,fetter1975}. In the case of Ne~\cite{borg1990b} and for He at $T>4\,$K~\cite{borg2002}, provision has been made for the bubble to be only partially empty because of the larger thermal energy of the atoms, which can penetrate the bubble walls.

\begin{figure}[t!]
 \resizebox{1\columnwidth}{!}{%
   \includegraphics{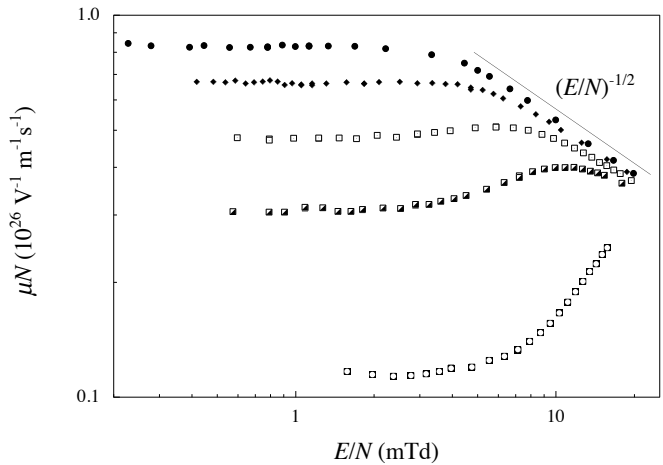}}
 \caption{$\mu N$ vs $E/N$ for Ne at $T\approx 46.5\,$K~\cite{borg1990b}. From top: $N\,$(nm$^{-3}$)$=9.02,\,9.56,\, 10.21,\,11.08,\, $ and $12.63.$ Line: 
cKT prediction for strong $E/N.$ The localization onset appears as a change of the 
$E/N$-dependence of $\mu N.$
 \label{fig:NeonLocalizatonOnset}}\end{figure}

In the simplified model~\cite{miyakawa1969} the excess free energy $\Delta F$ is given by
\begin{equation}\label{eq:DF}
\Delta F (T,N,R)= E_e+ E_p + E_s + E_v -V_0(N)
\end{equation}
where $E_e$ is the energy eigenvalue of the ground state of the electron localized in the void, $E_p$ is the polarization energy of the surrounding medium. $E_s$ and $E_v$ are the energy spent to create the bubble surface and the mechanical work spent to expand the bubble, respectively. $\Delta F$ is a function of $T,$ $N,$ and of the bubble radius $R.$ The minimization of  $\Delta F$ with respect to $R$ yields the most probable state with relative population $n_b=\exp{-\Delta F/k_\mathrm{B} T}.$   Once the radius $R$ of the bubble state is known, it is assumed that the localized electron mobility $\mu_B$ is given by the hydrodynamic Stokes formula
\begin{equation}\label{eq:stokes}
\mu_B=\frac{e}{6\pi\eta R }\end{equation} where $\eta$ is the gas viscosity.
The observed mobility is then a weighted average of the mobilities of quasifree $\mu_f$ and  localized $\mu_b$ electrons
\begin{equation}\label{eq:muave}
\mu= (1-n_b)\mu_f + \mu_b n_b
\end{equation}
with $ n_f+n_b=1.$
The quasifree electron mobility $\mu_f$ is computed according to the heuristic models previously described. The results for He at $T=26.1\,$K~\cite{borg2002} and Ne at $T=46.5\,$K~\cite{borg1990b} are reported in \figref{fig:MuvsNHeNeLocEgg}. The OFM accurately locates the threshold density beyond which localization takes place. From an experimental point of view, the threshold density can be located by inspecting how the field dependence of the mobility changes with changing density, as shown in~\figref{fig:NeonLocalizatonOnset} for Ne.
However, the prediction of the average mobility is quite poor. This conclusion is not surprising because only the most probable state is obtained by the OFM, whereas a whole distribution of radii is to be reasonably expected.

By contrast, a percolation model~\cite{eggarter1971} has been pro\-po\-sed, according to which electrons percolate through the islands and lakes of a disordered, fluctuating potential given by $V_0(N).$ Main goal of this model is to compute how the density fluctuations affect the electron density of state (DOS). In this way, the transition from fast to slow electron transport occurs in a smooth way (see~\figref{fig:MuvsNHeNeLocEgg}. 

Although the predictions of the percolation model appear to be in better agreement with the experiments, it has to be noted that there is an ambiguity related to the choice of the length $L$ over which the gas density has to be sampled in order to get fluctuations correctly. According to the original paper~\cite{eggarter1971}, $L\propto \lambda_T$ has to be chosen to get a nice agreement with the experiment in He for $T=4.2\,$K~\cite{levine1967,eggarter1971} and for $T=26.1\,$K~\cite{borg2002}. Unfortunately, to get the same agreement in Ne at $T=46.5\,$K, $L\propto \ell$ had to be used. This discrepancy arises from the fact that in He $\lambda_T\approx \ell,$ where as in Ne $\ell \gg \lambda_T.$ Apparently, the conclusion has to be drawn that the electrons sample the surrounding density by means of collisions, hence, over the distance of a mfp.

\section{Charge Collection Efficiency}
\label{sec:qc}
The electrons whose transport properties are to be studied must be injected into the gas. Photo-~\cite{tauchert1977,borg1986} or tunnel cathodes~\cite{silver1967,silver1970} have mainly been used to this goal.
Also the phenomenon of the charge injection may be studied in order to shed more light on the dynamics and energetics of electrons in a dense, disordered system. Upon injection, electrons are subjected to the two competing forces of the externally applied field $E$ that pulls them toward the anode and of the their own image that pushes them back to the cathode.  The combined potential energy is \begin{equation}\label{eq:vx}
V(x)=-eEx -\frac{1}{4}\frac{e^2}{4\pi\epsilon_r\epsilon_0 x}
\end{equation}
where $x$ is a coordinate from cathode to anode, $\epsilon_0$ is the vacuum permittivity and $\epsilon_r$ is the relative dielectric constant of the gas. Typically, $\epsilon_r\approx 1$ at the working densities.
The potential energy $V(x) $ has a maximum for $x=x_M= (e/16\pi \epsilon_0 E)^{1/2}. $ 

\begin{figure}[b!]
\resizebox{1\columnwidth}{!}{%
  \includegraphics{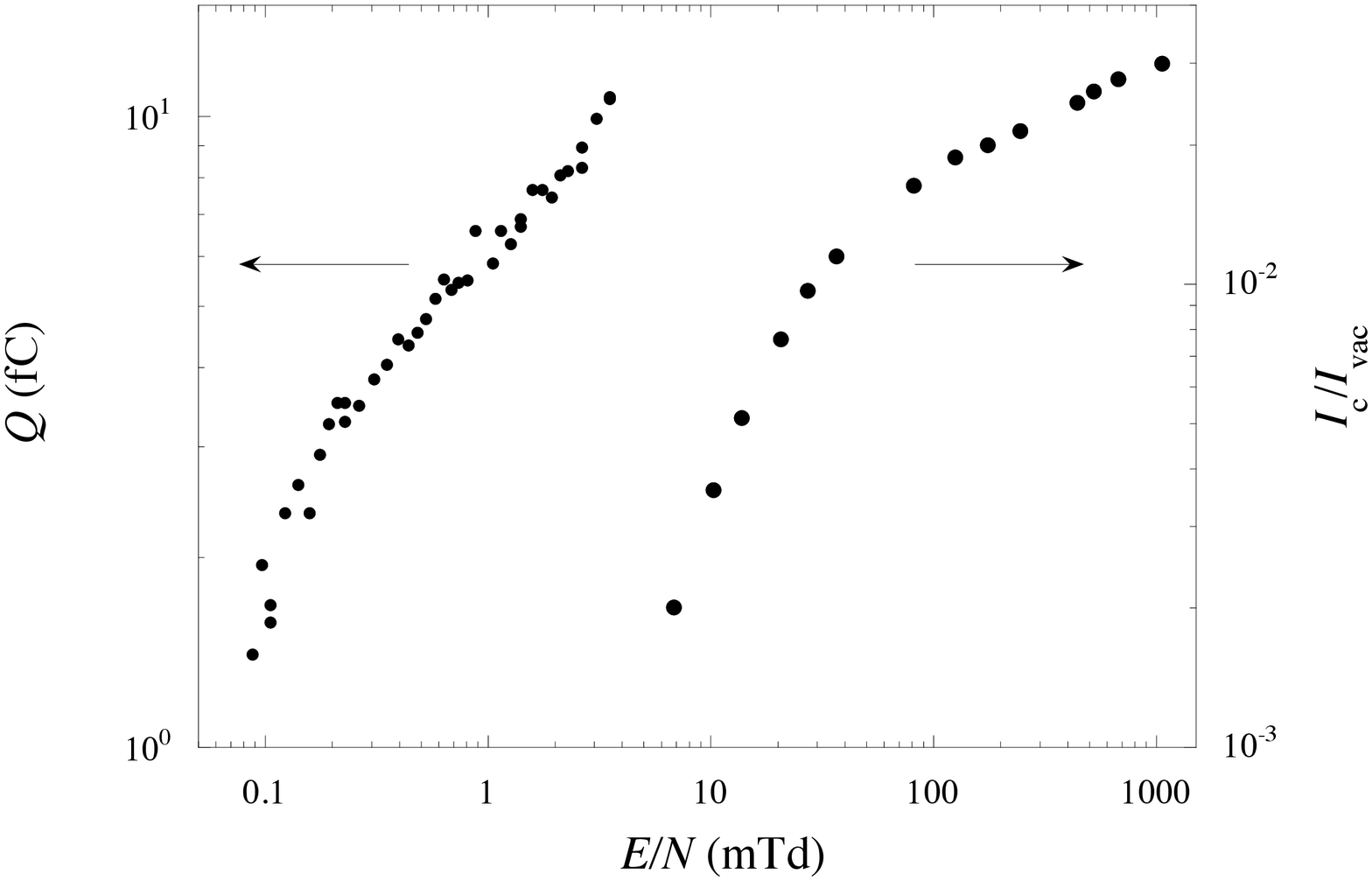}
}
\caption{$E/N$ dependence of the charge $Q$ collected in Ar gas for $T=152.7\,$K at $N=11.39\,$nm$^{-3}$ (left scale)~\cite{borg2013} and of the current injected into He gas by means of thin film emitters for $T=20.3\,$K at a $N=12.64\,$nm$^{-3}$ (right scale)~\cite{onn1971}. \label{fig:QargonIHelium}}\end{figure}

Typical results for the collected charge or current in He and Ar are shown in~\figref{fig:QargonIHelium}.
The two gases have quite different features. Namely, $V_0>0$ for He and $V_0<0$ for Ar, in addition to different energy dependence of the cross sections. These differences manifest themselves also in the field behavior of the charge or current collected at the anode. Note that in He $E/N$ values two orders of magnitude stronger than in Ar at nearly the same density are required to collect a significant current at the anode.

It has been shown~\cite{silver1970,onn1971,onn1969,smejtek1973} that the fraction of current or charge collected at the anode in a gas with respect to that collected in vacuo can be roughly expressed as
\begin{equation}
I=I_\mathrm{vac} \frac{A}{D}e^{-x_M/x_0}
\label{eq:iOnn}\end{equation}
$x_0$ is the electron thermalization length. $A$ describes the effect of the electron-gas surface barrier and $D=D(E,\ell)$ describes the back diffusion of unthermalized injection electrons~\cite{onn1969}.

\begin{figure}[t!]
\resizebox{1\columnwidth}{!}{%
  \includegraphics{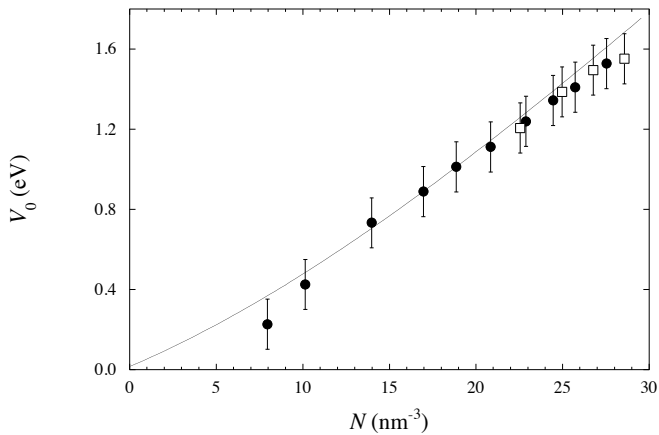}
}
\caption{Experimental determination of the density dependent energy shift $V_0$ in He at  $2.5\,\mbox{K}\,<T<20\,\mbox{K}$~\cite{broomall1976}. Closed points: gas. Open points: liquid. Line: Wigner-Seitz model~\cite{jortner1965}.\label{fig:BroomallV0He}}\end{figure}

In He, the large and constant cross section leads to quite short thermalization length and electrons are typically thermalized at a distance $x<x_M$ from the cathode. In this situation the collected charge depends on the features of the barrier to the injection $V_0(N)>0.$ An accurate analysis of the dependence of the collected current as a function of $E,$ $N,$ and $T$ has led to the experimental determination of the density dependence of $V_0(N)$ shown in \figref{fig:BroomallV0He}~\cite{broomall1976}.
The experimental determination of $V_0 (N)$ is in very good agreement with the theoretical predictions of the Wigner-Seitz model~\cite{miyakawa1969,jortner1965}.

On the contrary, the cross section in Ar leads to very long thermalization distances. In this situation, the backdiffusion to the cathode can be studied. There is only one, very old,  available model to treat backdiffusion in a dilute gas ~\cite{YB} which is quite successful at predicting the field dependence of the charge or current collected at the anode.
This model allows the determination of the cross section from the current data. Early attempts at determining the density dependence of $\sigma_\mathrm{mt}$ gave preliminary, not correct results because the influence of MS on the mobility was not yet clear at that time~\cite{smejtek1973}.

\begin{figure}[t!]
\resizebox{1\columnwidth}{!}{%
  \includegraphics{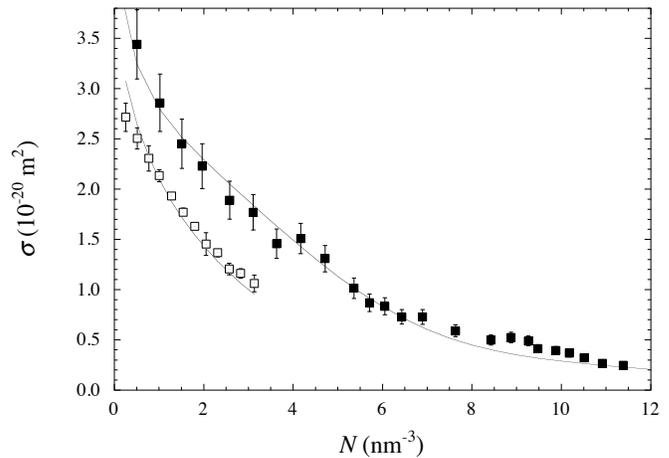}
}
\caption{Density dependence of the momentum transfer scattering cross section from charge collection efficiency experiment in Ar gas for $T=152.7\, $K (closed squares) and for $T=199.7\, $K (open squares)~\cite{borg2013}. Lines: heuristic model~\cite{BSL}.\label{fig:SigmaQT153and200}}\end{figure}

More recently~\cite{borg2013,borg2011}, the charge collection efficiency data in dense Ar gas have been analyzed by taking into account MS effects within the same heuristic model for the drift mobility described previously. In \figref{fig:SigmaQT153and200} the agreement between the values of the momentum transfer scattering cross section determined from the charge collection efficiency is compared with the density dependent cross section given by~\eqnref{eq:sigmastar}. The agreement is very good, thereby confirming once more the important effect of MS and also the validity of its treatment within the heuristic model.

\section{Resonant Electron Attachment to Oxygen}
\label{sec:ra}
In order to be able to carry out accurate measurements of drift mobility in gases, the amount of electron attaching molecular impurities, among which the most abundant is O$_2,$ must be reduced to the parts-per-billion (ppb) level. The purest commercially available gases have impurity content of the order of tens of parts-per-million (ppm). Oxygen has an electron affinity $E_A\approx 0.46\,$eV and rapidly scavenges the drifting electrons leading to the formation of very slow O$_2^-$ ions. 

\noindent 
This electron attachment process that may be lethal for the measurements of electron flight-of-time is, however, a very important and useful phenomenon in the domain of electrical insulation in high-voltage equipments. Actually, electron attaching species as, for instance, SF$_6$, are used as gaseous insulators because they quench any incipient discharge. For this reason, the investigation of the features of the attachment process is very important.

The attachment process at low density is a two-step resonant process~\cite{bloch1935}: the O$_2$ molecule in its vibrational ground state attaches the electron giving origin to the negative ion in a vibrationally excited state~\cite{boness1970}. Subsequent collisions with the gas atoms $M$ carry away the excess energy $\epsilon$ and stabilize the anion according to the reaction scheme~\cite{mark1985,matejcik1996}
\begin{eqnarray*}
\mathrm{O}_2 (X^3\Sigma_g^-;v''=0)+ e &\rightarrow &\mathrm{O}_2^{-\star} (X^2\Pi_g;v'\ge 4)
\\
 \mathrm{O}_2^{-\star} (X^2\Pi_g;v'\ge 4)+M &\rightarrow& \mathrm{O}_2^{-} (X^2\Pi_g;v'\le 4)+M +\epsilon
\end{eqnarray*}
It is a resonant process because only electrons with the proper energy ($E_R\approx 90\,$meV for $v''=4$) can produce the ion in the first available vibrationally excited level. The attachment process can thus be exploited to probe the electron energy distribution function $g(\epsilon)$ at fixed energy to investigate how $g$ may change with density. Actually, it can be shown that the reduced attachment frequency is proportional to the value of the electron energy distribution function at the resonance energy $ \nu_A/N\propto g(E_R).$

\begin{figure}[t!]
\resizebox{1\columnwidth}{!}{%
  \includegraphics{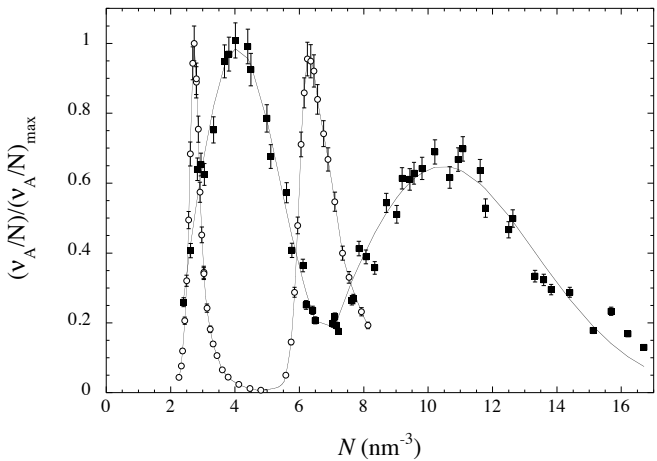}
}
\caption{ Normalized reduced attachment frequency $\nu_A /N$ vs $N.$ Open symbols: Helium gas at $T=54.5\,$K~\cite{borg1997}. Closed symbols: Ne gas at $T=46.5\,$K~\cite{borg1991}.  The lines are only a guide for the eye.\label{fig:NuAHeNe}}\end{figure}

According to cKT, $\nu_A /N$ should not depend on $N.$ Actually, early measurements in He at $T=77\,$K have shown the existence of a peak of $\nu_A/N$ for $N\approx 3\,$nm$^{-3}.$ More recent measurements, both in He at lower $T$~\cite{borg1997} and in Ne at $T=46.5\,$K~\cite{borg1991}, close to the critical temperature $T_c=44.4\,$K, have confirmed the older results and have also shown the presence of a second attachment peak at much higher $N$ in both gases. The results of the reduced attachment frequency $\nu_A /N$ are shown in \figref{fig:NuAHeNe}.
In both gases, the two peaks are clearly shown. The second peak at higher $N$ is due to the attachment of electrons to the second available vibrational level of the oxygen ion, namely that with $v^{\prime\prime}=5, $ with higher resonance energy $E_R\approx 215\,$meV.  

The explanation of the presence of such peaks in $\nu_A$ can be readily understood once the concept of a density dependent shift $V_0 (N)$ has been clearly assessed by the mobility measurements. As $N$ is increased, the electron energy distribution function $g$ is shifted by $V_0 (N)$. In He and Ne, owing to their quite small atomic polarizability, $V_0(N)\approx E_k(N)>0.$
The attachment process probes $g$ at the constant energy $E_R\approx 90\,$meV (or 215 $\,$meV for the second peak). Thus, the shape of the $\nu_A (N)/N$ curve is a replica (in density) of the electron energy distribution function.

The position $N_M$ of the peak is such that $V_0(N_M)\approx E_R.$
The second peak is simply due to the attachment of electrons to the next available vibrational level $v''=5$ of O$_2^-, $ for which $E_R\approx 215\,$meV. The densities of the peaks in Ne are larger than in He only because the energy shift $V_0$ is smaller, owing to the smaller scattering length.
In a certain sense, the attachment process can can be used to carry out molecular spectroscopy in a dense gas of repulsive scatterers.

In Ar, there is experimental evidence that there are no attachment peaks. This lack of structure of $\nu_A /N$ in Ar is due to the fact that both $V_0$ and the ion polarization energy are negative and nearly equal to each other so that the resonance conditions are never met~\cite{borg1997}.

\begin{figure}[t!]\vglue .75 cm
\resizebox{1\columnwidth}{!}{%
 \includegraphics{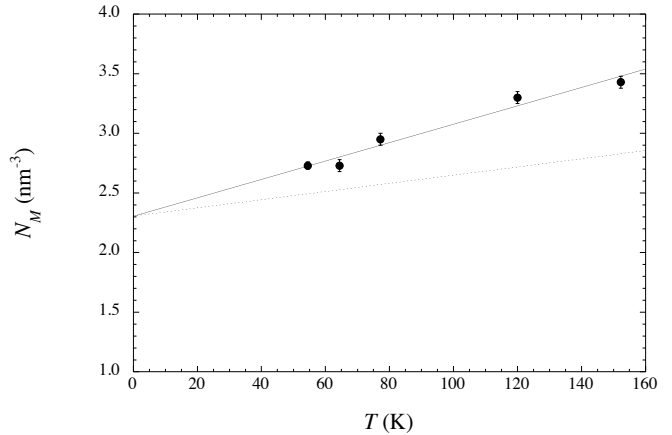}
}
\caption{$N_M$ vs $T$ in He gas~\cite{borg1997}. Solid line: linear fit to the data. Dashed line: ionic bubble model~\cite{volykhin1995}.\label{fig:NMdiTHelium}}\end{figure}

In He it has been possible to measure the first attachment peak as a function of $T$ in the range $54\,\mbox{K}\lesssim T\lesssim160\,\mbox{K}.$ The density of the peak $N_M$ is plotted as a function of $T$ in \figref{fig:NMdiTHelium}.
Na{\" i}vely, one would expect that $N_M$ should decrease with increasing $T$ because there is more thermal energy to be added to $V_0(N)$ in order that the mean electron energy equals $E_R.$ Actually, one has to consider that a small void is to be expanded around the ion as a consequence of the repulsion between the extra electron in the ion and the electronic clouds of the surrounding atoms that leads to the so called {\em ionic bubble}~\cite{volykhin1995}. Thus, the expansion work must be considered in the energy balance. By so doing, the dashed line in~\figref{fig:NMdiTHelium} is obtained that is in reasonable agreement with the experiment~\cite{borg1997}. The remaining discrepancy can be attributed to the neglect of surface tension work necessary to create the void interface.

Unfortunately, there still are unexplained features such as the huge width of the peaks in Ne for which a model for the computation of the distribution function in presence of strong density fluctuation should be developed. Actually, the measurements in Ne have been carried out quite close to the critical point where fluctuations, hence the compressibility, grow very large.

\section{Oxygen Ion Transport}
\label{sec:o2t}

The resonant electron attachment to O$_2$ molecular impurities can be exploited to investigate the transport properties of the O$_2^-$ ion in dense gases. 
Once more, dense gases are a no-man's-land between dilute gases and liquids. Whereas in the former the study of mobility is linked to the determination of the ion-atom interaction potential~\cite{mason}, in the latter the hydrodynamic Stokes' formula is traditionally used~\cite{schmidt}.

Early measurements of the drift mobility of O$_2^-$ anions in He gas at $T=77\,$K were carried out but well below $N_c$~\cite{bartels1975} in a density region in which the mobility can be estimated from the ion radius in the so-called Knudsen regime~\cite{khrapak1981}. 

\begin{figure}[t!]\vglue 1 cm
\resizebox{1\columnwidth}{!}{%
  \hglue 0.75 cm\includegraphics{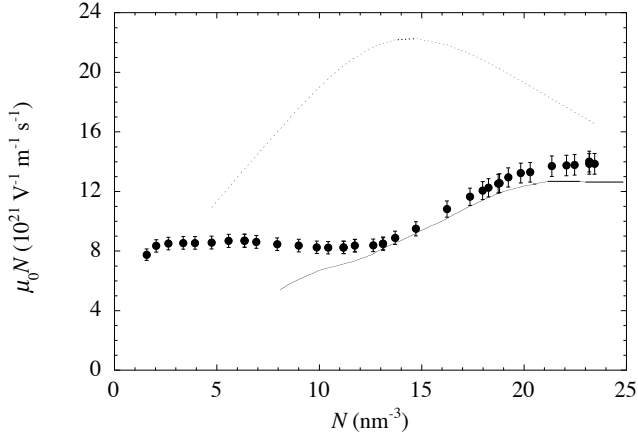}
}
\caption{$\mu_0 N$ vs $N$ for O$_{2}^{-}$ ions in Ne at $T=45\,$K~\cite{borg1993}. 
Dashed line: hydrodynamic Stokes formula for $R=4\,$\AA.
Solid line: the prediction of the hydrodynamic Stokes formula if the local $N$ and $\eta$ around the ion are modified by electrostriction and repulsive exchange forces according to the ionic bubble model~\cite{volykhin1995}.\label{fig:MuNO2NeonStokKhrap}}\end{figure}

\begin{figure}[t!]
\resizebox{1\columnwidth}{!}{%
  \includegraphics{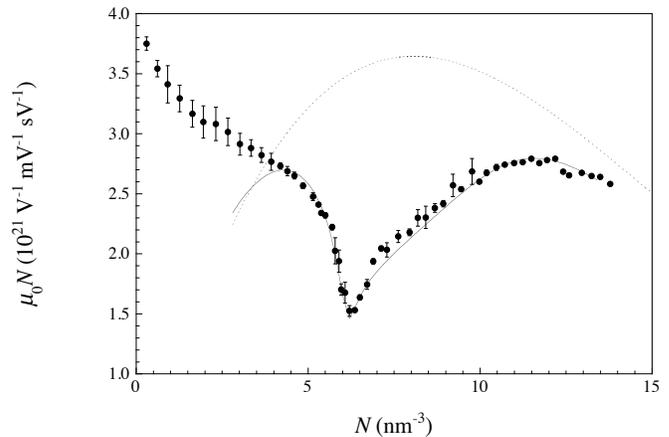}
}
\caption{$\mu_0 N$ vs $N$ for O$_{2}^{-}$ ions in Ar at $T=151.5\,$K~\cite{borg1997b}. 
Dashed line: hydrodynamic Stokes formula for $R\approx 6\,$\AA.
Solid line: prediction of the hydrodynamic Stokes formula if the local density and viscosity around the ion are modified by electrostriction. \label{fig:MuNO2ArgonStokesCPL}}\end{figure}

At much higher densities, the drift mobility of O$_2^-$ anions has been measured as a function of $N$ in  Ne gas at $T=45.0\,$K~\cite{borg1993}  and Ar gas at $151.5\,\mbox{K}\le T\le 157\,\mbox{K}$~\cite{borg1997b}. In both cases $T$ was close to the respective critical temperature ($T_c=44.4\,$K for Ne and $T=150.9\,$K for Ar) and the investigated density range includes the respective critical densities ($N_c=14.44\,$nm$^{-3}$ for Ne and $N_c=8.08\,$nm$^{-3}$ for Ar). In \figref{fig:MuNO2NeonStokKhrap} the experimental results of $\mu_0 N$ in Ne are shown, whereas the results in Ar are reported in \figref{fig:MuNO2ArgonStokesCPL}.
In the figures the dashed lines represent the hydrodynamic Stokes formula~\eqnref{eq:stokes}. The disagreement with the experimental data is evident. There is no way to choose a reasonable value of the ion hydrodynamic radius $R$ that reconciles theory and experiment. 

\begin{figure}[t!]
\resizebox{1\columnwidth}{!}{%
  \includegraphics{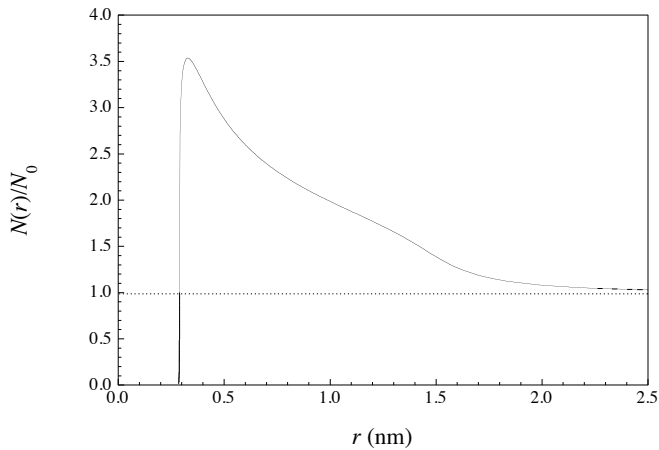}
}
\caption{Electrostriction-induced local density enhancement around the O$_2^-$ ion in dense Ar gas at $T=151.5\,$K for an unperturbed gas density $N_0= 6\,$nm$^{-3}$~\cite{borg1997b}. Qualitatively similar results have also been obtained for Ne gas by using a self-consistent field approximation  used in the theory of electron self-trapping~\cite{volykhin1995}.\label{fig:profiliArgonN60T153K}}\end{figure}

Hydrodynamics assumes that the density and viscosity of the gas around the ion are not affected by interaction with the ion itself. On the contrary, the ion interacts in two ways with surrounding fluid. First of all, the ionic charge polarizes the surrounding gas attracting it towards the ion thereby leading to a formation of a solvation shell. The density enhancement around the ion can be computed from the ion-solvent interaction potential within the electrostriction model~\cite{atkins1959}

\begin{equation}\label{eq:atkins}
-V(r)=K^2(N)\int\limits_{N}^{N(r)}\frac{1}{N^\prime}\left(\frac{\partial p}{\partial N^\prime}\right)_T\,\mathrm{d}N^\prime
\end{equation}
where $K$ is the gas dielectric constant and $p$ is the pressure. $N(r) $ is the local density value. Eq.~\ref{eq:atkins} can be easily solved for $N(r)$. 
A typical density profile for Ar at $T=151.5\,$K and for $N=6\,$nm$^{-3}$ is shown in~\figref{fig:profiliArgonN60T153K}
Also the viscosity $\eta$ is locally enhanced owing to its density dependence. 

In addition to electrostriction, there is also the repulsive exchange force between the extraelectron in the ion and the electronic clouds of the surrounding atoms. This interaction can be self-consistently treated within the ionic bubble model~\cite{volykhin1995} and leads to $N$ and $\eta$ profiles similar to those shown~\figref{fig:profiliArgonN60T153K}.

If the local enhancement of $N$ and $\eta$ is taken into account by using the $\eta$ value at the top of the density profile, the modified Stokes formula is now in  much better agreement with the data, as shown by the solid line in~\figref{fig:MuNO2NeonStokKhrap}.

The experiment in Ar is carried out even closer to the critical temperature than the Neon experiment. For this reason, the effective hydrodynamic radius has to be corrected by an amount proportional to the correlation length $\xi$~\cite{borg1997b} that can be easily related to the gas compressibility~\cite{croxton}. Again, as shown in~\figref{fig:MuNO2ArgonStokesCPL}, the modified Stokes formula (solid line) is in very nice agreement with the data at intermediate and high density. At lower density, hydrodynamics ceases to be valid.

It has finally to be noted that the drop in $\mu_0 N,$ which is small in Ne for $N\approx 13\,$nm$^{-3}$ and quite deep in Ar for $N\approx 6.2\,$nm$^{-3},$ occurs at a density smaller than $N_c$ in both gases. This observation can be easily rationalized by inspecting the density profile reported in~\figref{fig:profiliArgonN60T153K}. Owing to the local density enhancement around the ion because of electrostriction, the gas density takes on the critical value 
$N_c$ at some distance from the ion only if the density of the unperturbed gas is $N_0<N_c$. Thus, critical point effects related to the large compressibility occur in the gas near the ion only when the average density is smaller than $N_c.$ When the unperturbed gas density equals or exceeds $N_c$, then the gas is no longer so compressible and no critical point effects can be observed. 

Moreover, for extremely large densities, the modified- and original Stokes equations tend to converge to each other because the continuum approximation of an incompressible fluid applies also to a highly compressed gas.

\section{Conclusions}\label{sec:con}
In this paper the results of electron swarm experiments in dense rare gases have been reviewed. 
It has been shown that the experimental results on the electron drift mobility can be consistently described within a unique physical model, in which the MS scattering effects can be heuristically treated. In this way, the single scattering picture of classical Kinetic Theory can be retained, although the MS effects are manifestation of a many-body physics.

However, it has also been shown that the validity of this MS approach ceases at very high density in gases whose electron-atom interaction is dominated by short range repulsive exchange forces because thermodynamics induces the appearance of another phenomenon, localization, in which the electron ground state cannot be obtained as an adiabatic evolution of the quasifree state. The Optimum Fluctuation Model or the Percolation Model more or less successfully aim at the computation of the density of states of the electrons. In any case, a description of the average mobility must rely on the heuristic model for the calculation of the mobility of quasifree electrons.

 Also the electron charge or current collected at the anode in swarm experiments can be used to gather important pieces of information on the energetics and dynamics of the electrons. In He, in which the electron thermalization length is short, the dependence of the collected current allows to determine the barrier to electron injection, i.e., the ground state energy at the bottom of the conduction band, which is the main outcome of MS. By contrast, in Ar in which the electron thermalization length is much longer than in He, it is possible to investigate the properties of electron backdiffusion, from which an independent determination of the density dependence of the momentum transfer scattering cross section can be obtained. It has been shown that this determination is in very good agreement with that obtained by measuring the drift mobility.

It has also been shown that the concepts of MS have to be used to understand another phenomenon that takes places during the electron drift, i.e., the resonant attachment to O$_2$ molecular impurities. The existence of peaks in the reduced attachment frequency at given densities in Ne and He cannot be rationalized if the concept of a density dependent quantum shift of the mean electron energy were not been assessed in electron swarm experiments at high density.

Finally, the production of O$_2^-$ anions as a consequence of electron attachment to O$_2$ in a dense gas has allowed to study how the ion transport properties depend on how the fluid structure is locally modified by the interaction with the ion. This is not only of electrostatic origin but includes the short-range repulsive exchange forces between the extraelectron in the ion and the electronic shells of the surrounding atoms.

As a concluding remark it is possible to say that electron swarm experiments in dense rare gases have shed light on several important manybody effects that influence the dynamics and energetics of the electrons thereby bridging the gap between the dilute gas and the dense liquid.

\end{document}